\documentstyle[epsf]{binbook}

\newenvironment{abstract}{\section*{Abstract}\par}{}
\newcommand{\acknowledgements}{\par\vspace{\baselineskip}
	    \noindent {\it Acknowledgements\ }}

\def \gta {\mathrel{\vcenter
     {\hbox{$>$}\nointerlineskip\hbox{$\sim$}}}}
\def \m{\ifmmode M_\odot\else M$_\odot$\fi}
\def \l {\ifmmode L_\odot\else L$_\odot$\fi}
\def \etal {{\it et~al.}}
\def\deg{\ifmmode^\circ\else$^\circ$\fi}
\def \Teff{{\it T}\lower.5ex\hbox{\rm eff}}

\begin{document}
\setcounter{page}{0}
\pagenumbering{arabic}
\pagestyle{myheadings}

\author{%
         J. Craig WHEELER
         {\it Department of Astronomy}\\
         {\it University of Texas at Austin, Austin, TX 78712, USA}\\
         {\tt wheel@astro.as.utexas.edu}\\
         }
\chapter{Binary Evolution of Type Ia Supernovae}{J. Craig Wheeler}
\markboth{J. Craig Wheeler}{Binary Evolution of Type Ia Supernovae}
\label{cw}

\begin{abstract}
Observations of Type Ia supernovae (SN~Ia) combined with modeling of dynamics,
light curves and spectra continue to point to the difficult conclusion
that SN~Ia result from degenerate ignition in a carbon/oxygen white dwarf of
the Chandrasekhar mass.  Such a model accounts well for the ``normal"
SN~Ia and for the observed dispersion exemplified by the light curve
amplitude-shape relation.  The criterion that the white dwarfs involved
in this process grow to nearly 1.4 \m\ continues to provide a
great challenge to understanding the binary stellar evolution
involved.  The so-called ``supersoft X-ray sources" may provide an important
new population of SN~Ia progenitors.  Other techniques are being developed
that may constrain the binary evolution.
\end{abstract}

   \section{Introduction}
   \label{cwintro}

We {\it know} that Type Ia supernovae (SN~Ia) result from binary evolution.
The only caveats are that 1) there is no observational evidence for this
statement, and 2) there is no generally accepted theory for the
evolution.  What must this binary evolution accomplish?  This is driven
by the current state of observations summarized in \S~\ref{cwobs}.
The next question is the physical mechanism of the explosion. This is
constrained by
current theory of the nature of the combustion mechanism as presented
in \S~\ref{cwtheory}.  These two areas are then brought together
by considering the types of observed binary systems that are viable
progenitor candidates: cataclysmic variables, symbiotic variables,
supersoft sources, double degenerates.  These candidates are discussed
in \S~\ref{cwprogen}.  The issue of particular observational
evidence and constraints that may help to close the link between the
theory and observation of SN~Ia and the theory and observation of binary
stellar evolution is given in \S~\ref{cwconstraints}. Conclusions
are given in \S~\ref{cwconcl}.  A nice review article covering some
of these same topics is given by Branch \etal\ (1995)

\section{Observations of Type Ia Supernovae}
\label{cwobs}

The long-suspected dispersion in light curve behavior in terms of
a correlation between the peak brightness and subsequent rate of
decline (Pskovskii 1977;
Branch 1981)
 has been beautifully and
substantially confirmed by recent systematic, high-quality, photoelectric
photometry (Phillips 1993; Hamuy, Phillips, and Maza 1994;
Maza \etal\ 1994).  This work shows that there
is a systematic amplitude/decline rate relation in the sense that
the brighter SN~Ia decline more slowly.  They are also bluer at maximum,
have a larger {\it minimum} velocity of calcium (Fisher \etal\ 1995)
and a pronounced secondary
infrared bump.  The dimmer extreme are fainter, redder, slower, and have no
secondary IR bump.  The prototype of the bright extreme is SN 1991T
(Filippenko \etal\ 1992a; Phillips \etal\ 1992).
This variety is observationally rare and hence must be intrinsically rare.
The prototype of the dimmer extreme is SN 1991bg
(Filippenko \etal\ 1992b; Leibundgut \etal\ 1993).  This type
is observationally rare, but since there are selection effects, it
may be intrinsically common.  Suntzeff (1995) has done Monte Carlo
calculations that indicate that perhaps 40\% of all SN~Ia could be
of this type.  It is presently somewhat controversial as to what one
means by a ``normal" SN~Ia and whether that category can be properly
defined.  It remains true that most observed SN~Ia are very homogeneous
in terms of both their light curve and spectral evolution and that
there is a {\it de facto} ``normal" SN~Ia against which one measures
the departures that constitute the dispersion of properties.  The
distribution of properties of SN~Ia may not be represented by a single
parameter
family.  The current state of the art is to seek such a family
(Reiss, Press, and Kirshner 1995; Hamuy \etal\ 1995).  Rather, there
may be a true dispersion around the mean amplitude/decline rate correlation.
The recent SN~Ia 1994D seems to be anonymously bright for its decline
rate (H\"oflich 1995).

As a cosmological aside, it is worth noting that, while
accounting for the mean light curve shape correlation has given
encouraging results in terms of measuring the distance scale using
SN~Ia as a ``calibrated candle," one must understand the dispersion
around the mean behavior and its associated Malmquist bias before
one can usefully use SN~Ia at redshifts of Z $\sim$ 1/2 - 1 to
constrain q$_0$, $\Omega_0$, and the cosmological constant, $\Lambda_0$.

There may be important clues to the nature of the progenitor systems
of SN~Ia in the correlation with the properties of the host galaxy.
The bright, slow-declining events seem to be missing in early type
E and S0 galaxies (Hamuy \etal\ 1996b).  There is a related correlation
of the supernova luminosity, the light curve decline rate, and the ejecta
velocity as measured by the main Si II line at early times (Branch and
van den Bergh 1993) and
the red edge of the Ca II H and K absorption lines at all epochs
(Fisher \etal\ 1995).  The implication is that there is at least
a loose correlation of explosion strength and galaxy type.  There
may be an even tighter correlation of explosion strength with
the age of the system.  Spirals may blur this correlation since
they have a mix of young and old stellar populations.

The question of the age of the progenitor population of SN~Ia is
an old and honorable controversy.  The standard argument has
been that the presence of SN~Ia in elliptical galaxies which
have presumably long since exhausted their star formation is
evidence that SN~Ia derive from an old population,
with a time-delay set by the process of binary evolution that
triggers the explosion.  Oemler and Tinsly (1979) argued
that the enhanced presence of SN~Ia in star-forming irregular
galaxies suggested that some SN~Ia, at least, arose in young,
short-lived stellar systems.  Maza and van den Bergh (1976)
argued that unlike Type II (and Type Ib and Ic), SN~Ia in
spiral galaxies show no correlation with spiral arms, and hence
are not especially young.  More recent studies of that question
(Bartunov, Tsvetkov and Filimonova 1994; Della Valle and Livio 1994; McMillan
and Ciardullo 1995) have suggested that SN~Ia may indeed be
correlated with spiral arms in spiral galaxies, but not
with H II regions as are the Type II and Type Ib/c.  The time
for stars to diffuse out of the potential of the spiral arms
is about 10$^8$ years (Biermann and Tinsley 1974).  That has long
been regarded as a lower limit to the age of SN~Ia progenitors,
based on the results of Maza and van den Bergh.  If SN~Ia
are, rather, associated with spiral arms, then they
must have an age of less than 10$^8$ years.  Such a result would play
havoc with reigning ideas on nucleasynthesis of iron and oxygen where the
down turn in the ratio of iron to oxygen at a ratio of iron to hydrogen at
a tenth of the solar value is supposed to be the onset of iron production by
SN~Ia rather early in the history of the Galactic disk (Wheeler, Sneden,
and Truran 1989: Ruiz-Lapuente, Burkert, and Canal 1995).
The implication of the constraint
that SN~Ia not be associated with H II regions is not so clear.
They could be born from stars that do not generate H II regions
or they could live long enough to drift away from their birth sites.  Either
would
seem to require an age in excess of 10$^7$ years and to leave
little room below the upper limit suggested by a spiral arm
population.  Another way of addressing this issue was given
by van den Bergh (1990).  Van den Bergh pointed out that if
SN~Ia come selectively from old systems where mass transfer
occurs from a companion giant, then they should be associated with
the old, red, component of both spiral and elliptical galaxies.
Van den Bergh showed, rather, that the rate of SN~Ia per unit
H band luminosity is much higher in spirals than ellipticals.
This suggests that there is not a common population of old
progenitors, but that there is some source of SN~Ia in
spirals in excess of that which can be accounted by the old
population.  There must be an appreciable contribution to SN~Ia in
spirals from young systems.

The net result of these developments is to suggest that at least
some SN~Ia arise in younger systems and probably give rise,
on average, to brighter, higher luminosity, slower declining
supernovae.  The question of whether there
is still some on-going star formation even in elliptical galaxies,
especially those that are hosts of SN~Ia is another interesting
complication.

\section{Theory of the Explosion Mechanism of Type Ia Supernovae}
\label{cwtheory}

The most stringent observational constraint on theoretical models
of supernovae is the spectral evolution.  As argued earlier, although
there is a clearly established dispersion in properties of SN~Ia,
there are events that conform to a well-defined spectral evolution
that must be understood as a framework in which to study the
dispersion.  Representative ``normal" SN~Ia in terms of their
spectral evolution (which is not to say that they do not have
some dispersion in light curve behavior) are SN 1972E, SN 1981B,
SN 1989B and SN 1992A.  An important goal is to develop a model
of dynamics and radiative transfer that accounts for this observed
spectral evolution.

For the best current radiative transfer and opacities, the velocity
and composition distribution of model W7 of Nomoto, Thielemann and
Yokoi (1984) remains the standard to beat to reproduce the
spectral evolution from near maximum to about 30 days later
at the start of the exponential decline.  Although there is
progress in developing nonLTE radiative transfer models, LTE models
are remarkably successful near maximum light with an appropriately detailed
treatment of opacity.  The opacity is the critical issue.  In the
supernova ambiance, it is composed of bound-free, free-free
and resonant line scattering.  The latter is the most difficult
and controversial.  The line scattering involves ``expansion opacity"
effects by which one envisions a photon emitted
from one type atom with one velocity being Doppler shifted in the
expanding atmosphere to be in resonance with another type of atom entirely
at a different position in the flow.  From this point of view,
the opacity can be strongly affected by the many weak lines that
would not be important in a static atnosphere, but which can provide
a forest of lines abetted by the Doppler shift that greatly
increases the net opacity.

One technique that has met with considerable success is that of
Harkness (1991ab; Wheeler, Swartz, and Harkness 1993; Wheeler,
Harkness, Khokhlov, and H\"oflich 1995) in which the population
levels are treated in LTE, but the photon distribution is treated
in full generality.  The structure is a ``snapshot" at a given
instant of time, so it not a fully consistent radiation hydrodynamics
calculation.  The opacity is computed in the co-moving frame where
it has the same value as in the laboratory.  The emitted spectrum
is computed by a Lorentz transformation to the rest frame of
the observer.  In his most recent models, Harkness (private communication)
computes the opacity of 15,000 strong resonance lines explicitly,
including resolution of the line profile, with 200,000 frequency
points.  He also includes 42,000,000 weaker lines with suitable
frequency averaging techniques.  One of the most important features
of the resulting spectrum is the strong ultraviolet deficit
that characterizes SN~Ia shortward of the Ca H and K lines at
about 4000 \AA.  Harkness (1986) showed that this is due to the resonant
scattering of many lines of iron-peak species.  His most recent
calculations show spectacular agreement with both the optical and
UV spectra of SN 1992A near maximum light for model W7.  The same
model also gives a quite good representation of the subsequent
spectral evolution of events like SN 1989B (Wells \etal\ 1994),
in particular the disappearance of the strong lines of intermediate
mass elements, Si, Mg, S, near maximum to be replaced by
growing iron lines.  This transition happens too quickly if
the opacity is not large enough, but the phase, sometime after
maximum, is about right in the current calculations based on model W7.

There is little question that the light curves and spectra of SN~Ia
demand the thermonuclear explosion and total disruption
of a carbon/oxygen white dwarf.  There is
some controversy over whether that white dwarf must be of the
Chandrasekhar mass or can or must be of some lower mass.  The latter
suggestion has been driven at least qualitatively by the recognition
that there is a dispersion of SN~Ia properties.  Interestingly, the current
preponderance of evidence suggests that all SN~Ia are of the
Chandrasekhar mass and that such models show more promise for
accounting for the dispersion in properties than do the sub-Chandrasekhar
mass models.

Sub-Chandrasekhar mass progenitors would, in principle,
be more abundant and easier to grow in a binary system.  They
also intrinsically avoid problems with excess electron capture
and neutronization of the ejecta that plague the more massive
models.  The problem is that they
do not match the direct observations well.  In these models, a
substantial degenerate helium shell accumulates on top of a
carbon/oxygen core (Woosley and Weaver 1994;
Livne and Arnett 1995).
The helium ignites and detonates first and can detonate the carbon
either directly or by generating an inward compression wave that
ignites the carbon in the interior of the core (Livne and
Glasner 1991).
The helium burns to
nuclear statistical equilibrium, but since the carbon and oxygen
are at relatively low densities, those fuels burn to produce
only intermediate mass elements in the outer portions of the underlying
C/O core.  The problems
with these models are that with the hot outer radioactive shell
of burned helium, they are too hot and too blue (H\"oflich
and Khokhlov 1995).
Although the models produce intermediate
mass elements, they tend to be at too small a velocity.  One of the
one-dimensional models of
Woosley and Weaver (1994) gives silicon at velocities up to 13,500
km s$^{-1}$, but more  typical maximum values are less than about
12,000 km s$^{-1}$.  H\"oflich and Khokhlov get similar results.
Some of the two-dimensional models of Livne and Arnett give Si up
to velocities of 15,000 km s$^{-1}$, but the mass fraction still
falls below 0.1 at velocities of about 13,000 km s$^{-1}$.  By
contrast, model W7 has Si at mass fractions exceeding
$\sim$0.25 up to 15,000 km s$^{-1}$. Ironically, although
sub-Chandrasekhar models can account for the luminosity of
``normal" SN~Ia, they do not account easily for the extremely subluminous
variety (note that Woosley and Weaver and Livne and Arnett
get luminosities roughly a factor of two less than H\"oflich
and Khokhlov; this may be due to a crude treatment of the
$\gamma$-ray deposition in the outer layers of expanding detonated
helium in the former).

Although the principle arguments against this class of
sub-Chandrasekhar mass models are observational,
there are also many remaining theoretical issues.  This model
demands off-center ignition and so it is clear that one-dimensional
calculations may be misleading.  In two-dimensional models,
the ignition at the helium-carbon/oxygen interface may be a dud,
so there may be a requirement that there already be a Chapman-Jouguet
detonation in the helium that may or may not occur, but requires
ignition within the helium layer, not right at the boundary.
This ignition leading to a detonation may be prevented by convection
in the ignited layer.  Even without detonation at the boundary, there
may be a shock wave from helium ignition that converges
in the carbon/oxygen core to ignite a detonation there, but then
the behavior in three dimensions may become critical.

On the other hand, some classes of Chandrasekhar mass models seem
to work quite well to account for the current observations.
Central carbon detonation seems inconsistent with the observations
since much of the star is processed to iron peak and there is
insufficient production of intermediate mass elements at moderate
velocities.  Even then, carbon detonations are unstable at some scale
(Khokhlov 1993; Boisseau 1995) and this may lead to less complete
burning than in one-dimensional models.  Central carbon deflagration,
in which there is a precursor shock and subsonic burning gives qualitatively
reasonable conditions.  This is because the model pre-expands so the
burning occurs at low density, does not go all the way to nuclear
statistical equilibrium in a substantial portion of the matter, and
hence leaves intermediate mass elements.  Pure carbon deflagration
models with self-consistent physics do not agree quantitatively
with the observations, but tend to leave the intermediate mass
elements in too narrow a shell in velocity space.  One-dimensional
models that make a transition from deflagration to detonation
in a prescribed way, the so-called delayed detonation models
can give a more reasonable composition/velocity
profiles (Khokhlov 1991a).  A very interesting variation on this
theme is the class of pulsating delayed detonation models
(Khokhlov 1991b; Khokhlov, M\"uller,
and H\"oflich 1993; Arnett and Livne 1994).
In these models, the first deflagration stage
causes the white dwarf to expand, but insufficient energy is
invested to eject mass.  The carbon re-ignites and (by assumption)
detonates when it again is compressed to densities above which it
can burn.  This class of models is plausible in the sense that
it gives a natural explanation of why the detonation ignition
occurs at low density. The carbon is re-ignited as it increases density
and burns as soon as it crosses the low density threshold.
The nature of the deflagration to detonation transition is
a challenging problem, unsolved in general even for most terrestrial
problems.  The transition may require destrucyion of the flame surface by
turbulence and a substantial mixing of burned and unburned material
(Khokhlov, Oran, and Wheeler 1996).  This is expected to occur
naturally in the context of the pulsating models, but must be demonstrated.

The current class of delayed detonation and pulsating delayed
detonation models rely on a free parameter, the density at
which the transition to detonation is made.  Interestingly,
within this freedom, these Chandrasekhar mass models
can reproduce the observed dispersion in light curve properties.
The key is that the bulk of the energy to drive the explosion
comes by burning from carbon to silicon.  Very little energy
is liberated in burning the rest of the way from silicon to
iron peak.  All the delayed detonation models burn substantial
portions of the mass to at least silicon and other intermediate
mass elements and thus provide ample kinetic energy.  These models
can, however, give a range in nickel mass.  This breaks the
direct tie between explosion energy and nickel mass which
has in the past been used to constrain the distance scale
(Sutherland and Wheeler 1984; Arnett, Branch, and Wheeler 1985).
Rather, some models with ample energy give small nickel mass and
hence less radioative heating.  Not only is the luminosity less,
but at smaller temperature the opacity is less, and this
serves to give a steeper decline from maximum and redder colors, just the
behavior
required for the light curve amplitude-decline rate relation.
The ensemble of such models provide a reasonable agreement
with the ensemble of observations (Hamuy \etal\ 1995) and
models can be found which match individual supernovae with
reasonable success (H\"oflich, Khokhlov, and Wheeler 1995;
H\"oflich and Khokhlov 1995).

One of the most interesting
success stories of this class of models is that one pulsating
delayed detonation model was identified as reproducing the
multi-band light curves of the especially dim event SN 1991bg
(H\"oflich, Khokhlov, and Wheeler 1995).  This model, with
no further adjustment, gave a superb fit to the maximum light
spectrum (Wheeler, Harkness, Khokhlov, and H\"oflich 1995).
This provides strong encouragement to think that the
class of Chandrasekhar-mass models represents the best
current paradigm to account for the observations of SN~Ia.

This is not to say that there are not still important problems
to be addressed to better understand the physics of these
models and to refine the connection to observations.  In their
own way, the deflagration models are just as dependent on
three-dimensional physics as the sub-Chandrasekhar mass models.
There is much yet to be done in the
venue of three-dimensional calculations.  Khokhlov (1995) has
shown with full three-dimensional calculations in a rectangular
box that the deflagration flame in a white dwarf propagates
according to a scaling law which is basically the Rayleigh-Taylor
timescale on the largest scales.  The speed of the front is independent
of the laminar flame speed on small scales.  This is because the
flame wrinkles to incorporate the rate of mass ingested as
dictated on the largest scales.  For small laminar flame speeds
the flame front is more wrinkled, giving a larger effective
area.  For larger laminar flame speeds, but the same macroscopic
conditions and hence the same global flame speed, the flame is
less wrinkled with a smaller effective area on the scale of the
laminar flame.  For proper choices of the fuel, this scaling
law is testable in laboratory apparatus (Khokhlov, Oran, and Wheeler 1995).

Khokhlov (1995) also shows that this scaling law allows for the
three-dimensional computation of the full white dwarf.  One must
only resolve the largest scales where buoyancy, gravity, and
geometry set the flame speed.  At smaller scales, the scaling law
dictates how fast the flame propagates.  The result of these
calculations is that the flame is quenched in the expansion.  The
spherical expansion damps the explosion, only a small mass, about
5\%, is burned, and the star expands without exploding.  This is
very consistent with the class of pulsating  delayed detonation
models.  In order to make a successful explosion, there
must be a deflagration to detonation transition when the star
recompresses.  The pulsation phase may naturally lead to an
effective turbulent destruction of the flame front, a mixing of the
hot products of burning with the unburned fuel and the subsequent
production of a detonation by the
Ze'ldovich mechanism (Zel'dovich \etal\ 1969, 1970) in the recollapse phase.
In this process, ignition occurs in a composition and
temperature gradient leading to a shock of accumulating strength
and eventually to a detonation (Khokhlov, Oran, and Wheeler, 1996).
Khokhlov's three-dimensional calculations ``remember" the initial
conditions of the ignition.  This is unfortunate from the computational
point of view, but may be the key to understanding the dispersion
of properties of SN~Ia despite the essentially fixed mass of the
progenitor.

\section{Observed Binary Systems as Possible Progenitor Candidates}
\label{cwprogen}

The results of the previous sections strongly point to the conclusion
that SN~Ia do arise in carbon/oxygen white dwarfs of the Chandrasekhar
mass.  This conclusion puts great demands on the requisite
binary stellar evolution that leads to this outcome.

The simple model of SN~Ia ``known" to most astronomers
is that they arise in a binary
system when a red-giant companion to a previously formed
white dwarf finally evolves and transfers mass to the white dwarf,
leading it to catastrophic ignition.  The problems with this simple
model are legion and discussed thoroughly in the literature
(Iben and Tutukov 1984, Paczy\'nski 1985).
Briefly, the problems are that at low accretion rate white dwarfs
undergo nova explosions which, from the evidence of ejecta enrichment,
actually eject some of the inner core, reducing the mass of the white
dwarf despite the prior accretion.  At somewhat higher accretion rates,
a degenerate shell of helium is predicted to accumulate. This configuration
is the basis of the sub-Chandrasekhar models discussed above.  The
main problem is that nature does not seem to do this in the observed
sample of SN~Ia.  At higher accretion rates, both H and He can
burn in a quasi-stationary, non-degenerate way, but the requirement
is that the resulting progenitors must be very bright, and it has
not been clear that any such progenitor systems have been observed.
Finally for even higher accretion rates, the hydrogen that is
transferred from the companion is argued to accumulate as an envelope
around the white dwarf, vitiating it as a candidate SN~Ia.  The result
of this type of analysis, put overly simply, is that all accretion
rates of hydrogen from zero to infinity are ruled out by various
constraints.  It was such a conclusion that gave great impetus to
the investigation of double degenerate evolution and attempts
to discover white dwarf binaries.

Revisionism has reached the
stage when the problems are reviewed with some hope, perhaps,
of re-establishing something like the simple model (Wheeler 1991, 1992).
The revisionist approach is to go back over all those arguments and
see if there are flaws in any of them.  As one might expect,
there are some loose threads at each stage, one of which might
turn out to be devastating.

The low accretion rate regime is
relevant to classical cataclysmic variables.  There is still
considerable uncertainty in the census of such systems and
hence of the total reservoir of candidate SN~Ia progenitors.
One of the issues that has been abandoned in its original extreme
form is the subject of hibernation.  It is not clear that this
issue is dead when there are still questions of a population
of dim CV-like systems (Shara \etal\ 1993; Howell, Szkody, and Cannizzo 1995).

The argument that novae are antithetical to SN~Ia remains very
important, and probably true, but worth close examination.  Continuing
calculations of nova models have resulted in a shift to higher
mass accretion rates of the ``nova/dud" line (Wheeler 1991)
that separates
explosions from thermal flashes in the accretion rate - white dwarf
mass plane.  The compilation of models presented in Livio
and Truran (1992) suggest a nova/dud line about two orders of
magnitude higher in accretion rate than that invoked by Wheeler (1991)
is his discussion of these topics.  One interesting aspect of this,
not directly relevant to the current topic, is that this shift
automatically solves the problem of how nova disks can be stable
to the dwarf nova-like disk instability if the nova models demand
a low accretion rate (Wheeler 1991).  With the revised models, the novae can
readily
fall in a regime where they have an accretion rate low enough
to trigger a nova explosion, but still high enough to heat,
ionize, and stabilize the disk.

There is probably still much
more to be learned from nova models.  For one thing, most of the
current models are one-dimensional and novae, representing off-center
ignition, are inherently multi-dimensional.  In preliminary
calculations, Livne (1995) finds that an initial convective
phase results in dredging up of core material.  The convection
brings in fuel and the flame does not propagate.  The
burning is not smooth, but occurs in oscillations.  The resulting
convective velocities are about ten times those deduced from
one-dimensional calculations, but they remain subsonic.  The
affect of this on the outcome over a broad range of conditions,
and the issue of whether three-dimensions, with  qualitatively
different turbulent energy cascade than in two-dimensions, will substantially
alter the conclusions needs to be investigated.  Another possibly
critical issue is the glabal propagation of the burning.  Study
of the propagation of nuclear burning fronts in the context of
X-ray bursts on neutron stars shows that the ignition and glabal
flame propagation can be very complex, especially in the regime
where there only a few nuclear ``fires" propagating at laminar
flame speeds (Bildsten 1993, 1995).
No parallel multi-dimensional studies have been done in
the context of nova ignition.

Another reason to be cautious about current
one-dimensional nova models, is that they apparently made erroneous
predictions concerning Nova Cygni 1975.  The nova models predict
that when the mass accretion rate goes up, the explosion is quenched
by lifting the degeneracy in the hydrogen shell.  It was argued that
a magnetic field would concentrate the accretion at the poles,
effectively enhancing the accretion rate per unit area and
quenching the explosion (Livio 1983). On this basis, it was predicted that
the white dwarf in Nova Cygni could not be magnetic.  It was
subsequently found to contain a polarized, magnetized white dwarf
(Stockman, Schmidt, and Lamb 1988; Livio, Shankar, and Truran 1988).

If novae do remove core mass and hence reduce the mass of the white
dwarf, there is still an interesting question of the secular
evolution.  Wheeler (1991) discussed the question of whether
novae would peel mass off the white dwarf until it reached the
nova/dud line where the nova process would halt.  It is still far
from clear that such systems could subsequently evolve in a way that was
conducive to the production of SN~Ia, but there are some interesting
possibilities.  One is the role of magnetic braking.  If magnetic
braking works in the standard way envisaged to account for
the CV period gap, then as angular momentum is lost, the orbital
period shrinks and the accretion rate goes down. This would result
in secular evolution along the nova/dud line to lower core mass,
the wrong direction for SN~Ia.  There are questions in the literature
as to whether magnetic braking is operative in the same way in
all systems.  Shafter, Cannizzo, and Wheeler (1989) point out
that standard magnetic braking laws seem to be inconsistent
with the ratio of dwarf novae to nova-like variables as a function
of orbital period. Warner (1991) has noted that magnetic braking can
only account for the period gap at a well-tuned transfer rate
and that the gap would be larger at other transfer rates for the
same orbital period.  Perhaps nature finds a way to provide that
accretion rate, but it is not obvious.  These caveats do not
mean that one can ignore magnetic braking, but there remains
a faint possibility that it does not apply in some cases and
that the secular evolution along the nova/dud line could be
toward higher masses. This might be the case if nuclear or thermal
evolution drives the mass loss as is suspected for the supersoft sources
(see below).  There are still immense problems with
growing a white dwarf all the way to the Chandrasekhar mass in this
way, including the forming and ejecting of a common envelope
even with non-dynamic, thermal  shell flashes (Livio \etal\ 1990).  There are
many
ways to use up the reservoir of the companion star before
the white dwarf grows sufficiently.

If nova explosions in classic cataclysmic variables provide too large
an impediment to SN I~a, there are still other types of binaries that
can be examined as potential progenitors.
There has been some interesting discussion recently of the question
of whether there are sufficient symbiotic variables to
account for the requisite number of SN~Ia progenitors.  Munari and Renzini
(1992) argued in favor of this possibility, but Kenyon \etal\ (1993)
concluded that it remains an unlikely possibility.

The question of whether SN~Ia could arise in merging double-degenerate
systems (Iben and Tutukov 1984; Webbink 1984) remains very interesting.
No systems have yet been identified with M$_{tot}$ $\gta$ M$_{Ch}$
(Robinson and Shafter 1987; Bragaglia, \etal\ 1990; Foss, Wade,
and Green 1991; Marsh, Dhillon, and Duck 1995),
but the question of whether upper limits might just allow sufficient
progenitor systems is still debated (Iben and Tutukov 1991).  The double
degenerate
scheme might have some trouble accommodating the growing evidence
for some fraction of the SN~Ia being from moderately young systems,
but this is not precluded.  Such systems might be required to
account for the odd, bright systems like SN 1991T which also had
spectral peculiarities.

One of the most interesting recent suggestions is that the supersoft
x-ray sources (SSS) may represent SN~Ia progenitors.  Recall that one of the
possible progenitor categories of SN~Ia consisted of systems in which the white
dwarf accreted at a rate such that H and He burned quiescently.  This
would avoid novae and helium detonation supernovae, but requires
very bright progenitor systems.  The counter argument was that
no such systems were observed.  A counter-counter argument was
that they might radiate in unobservable, highly absorbed bands.
Pioneering observations with the Einstein satellite (Long, Helfand
and Grebelsky 1981) and recent
discoveries with ROSAT (Tr\"umper \etal\ 1991; Hasinger 1994)
have revealed a set of sources that share
many of the expected characteristics of this class of progenitor
candidates.  These sources have luminosities in the range
L $\sim 10^{37} - 10^{38}$ erg s$^{-1}$ and effective temperatures
of \Teff\ $\sim 4\times10^5 K \sim 35$ eV.  There
are currently about 10 known sources in the LMC/SMC, 4 in the Galaxy,
and 15 in M31.  Di Stefano and Rappaport (1994), estimate a total
in each of these systems, after allowance for extinction, of about
 50, 1000, and 2500, respectively.  One of these systems is an old nova, GQ
Mus,
and two are symbiotic variables.  There are still questions of
whether some could be accreting neutron stars or black holes, so
the issue of the homogeneity of the sample and whether all are
potential SN~Ia progenitors is still significant.  These bright,
hot sources should ionize the ISM around them, leading to
substantial nebulae. Nine SSS have been observed
in the LMC/SMC and only one
such nebula has been detected (Remillard, Rappaport, and Macri 1995).
This is a key issue since the nebula might give a measure of the
mass of the white dwarf.

One of the most interesting interpretations of the SSS is that of
van den Heuvel \etal\ (1992), who proposed a system consisting
of a slightly evolved secondary star, M$_2$ $\sim1.3-2.7$ \m
$>$ M$_{wd}$.  Such a system undergoes unstable mass transfer
but on a thermal, not dynamical timescale.  The resulting transfer
rate is in the range $1-4\times10^{-7}$ \m\ yr$^{-1}$ which is
predicted to give steady burning of the H and He without shell
flashes.  This model automatically gives about the right luminosity,
the orbital period $\sim$ 1 day, and with the radius of the white
dwarf, \Teff~ in the range observed for the SSS.  The model
predicts that over the lifetime of the system, $\sim1-4\times10^9$ yr,
the white dwarf could accrete a few tenths of \m\ of matter. Although
van den Heuvel \etal\ did not explicitly mention the possibility, this
is all very encouraging for consideration of these systems as
progenitors of SN~Ia.

The SSS do not yet represent an open and shut case that the progenitors
of SN~Ia have been identified.  There is still a serious question
of whether there are sufficient of them to reproduce the observed rate
of SN~Ia, especially if the white dwarf must evolve to the
Chandrasekhar mass.  This is not a trivial issue, but involves many
model-dependent assumptions.  Several groups are now trying to
make population synthesis models to address these issues (Rappaport, Di Stefano
and Smith 1994; Yungleson \etal\ 1995). There are many
issues of physics that need to be considered, reconsidered, in
this context.  The issue of the exact dividing lines between
stable and unstable hydrogen and helium shell burning, even
at constant mass accretion rate, needs to be re-examined, perhaps first
in one dimension with current physics, but certainly in multiple
dimensions to look at the systematics of global flame
propagation in analogy to the neutron star case where there are
different regimes of laminar and convective burning (Bildsten 1992, 1995).
Another important issues is the actual time
variable mass accretion rate of evolving systems.  Many of the
past exploratory studies of possible SN~Ia progenitors have
simply explored parameter space with the transfer rate as a
free parameter.  The issue now is how the transfer rate evolves for
a realistic system.  If the transfer rate is too low the system
may undergo helium shell detonation. If it is too high the disk
may swell and block the X-rays or even cause the development of
an extended envelope that would entirely change the observational
character of the system.

Some of the SSS show a time variability
that may give a clue to their physical state.  RX J0513.9-6951 in the
LMC shows an X-ray variability on time scales of weeks with no
apparent variation in optical output (Pakull \etal\ 1993).  This is
much too fast to be associated with the shell-burning timescale associated
with the mass transfer which might be of order 100 yr (van den Heuvel
\etal\ 1992).  Pakull \etal\ suggest that the variation may indicate
a change in the accretion rate onto the white dwarf and associated
variation in the effective white dwarf radius and hence \Teff.

The need to focus systems in a narrowly defined mass transfer rate
to make the production of SN~Ia as efficient as possible may
require some feedback
process that actively regulates the accretion onto the white dwarf.
Van den Heuvel \etal\ (1992) discuss the possibility that if the
transfer rate gets too large the envelope around the white dwarf will
expand, thus choking off the transfer.
A process that is often overlooked in this context is what happens
when enough mass accumulates around the white dwarf that it can
form an extended, ``red-giant" envelope.  It is very likely that
this is impossible.  It is empirically well established that
most systems that form a red-giant envelope surrounding a degenerate
carbon/oxygen core eventually eject the envelope to form a
planetary nebula.  One of the most substantial theories in the
literature for this so-called ``superwind" process is the
catastrophic pulsational overstability that naturally occurs
driven by the luminosity of the double shell source surrounding
the degenerate core (Barkat and Tuchman 1980a,b and references therein).
If this process
is active (or whatever the physical mechanism behind the ejection
of planetary nebulae is!), then as the white dwarf grows in mass
it will very likely reach a state where sufficiently rapid accretion
to accumulate an envelope will result in the ignition of the
double shells and the ejection of that envelope.  This may
provide an upper limit to the actual accretion rate onto
the white dwarf which is automatically restricted to the upper
range above which an extended envelope tries to form. This
is just the range deduced for the SSS.  There are, of course,
still substantial questions of whether this process is so
wasteful of transferred matter that the reservoir on the secondary
star will be used up before the white dwarf reaches the point
of central carbon ignition.  It is interesting to note that the
observations of Pakull \etal\ (1993) may suggest that some
SSS, at least, do have P-Cygni features and hence associated
mass loss.

Livio and Truran (1992) discuss the possible connection of recurrent
novae to SN~Ia.  They argue that for recurrent novae to be
explained in terms of thermonuclear runaway, the white dwarf
mass must exceed 1.3 \m\ and the accretion rate $10^{-8}$ \m\ yr$^{-1}$.
They point out that these condition are very near those needed
for a SN~Ia explosion with a massive white dwarf still accreting
and growing. They do not offer an explanation of how the white
dwarf got that large mass in the first place and hence side-step
one of the critical issues.  Another issue that deserves
clarification is the interpretation of the nova/dud line.  Livio
and Truran note that some calculations give thermonuclear outbursts
at very high accretion rate if the white dwarf mass is sufficiently
large.  They discuss this as a strong deviation of the nova/dud
line up into a mass accretion rate regime relevant to the
SSS and the possible accumulation of an extended envelope (see below).
This is a legitimate interpretation in terms of the observed
properties of recurrent novae, but is not the only relevant factor
to the issue of the progenitors of SN~Ia.  The thermonuclear models for
recurrent novae at very high accretion rates do not, in fact, eject any mass.
Thus the critical nova/dud line for outburst is not the same as
the critical nova/dud line interpreted as the boundary for mass
ejection.  This is a complex issue, since even at more moderate
core masses accretion at a high enough rate to avoid a dynamical
nova event may nevertheless undergo a thermal shell flash
leading to a swelling of the white dwarf
envelope to form a common envelope around the secondary and hence
to mass ejection (Livio \etal\ 1990).  Nevertheless, if the nova/dud line has
any potential to affect the secular evolution of these systems,
then it is the mass ejection/no mass ejection version of the
nova/dud line that is relevant, not the flash/no flash version.
The former may intersect the Chandrasekhar mass limit at accretion
rates somewhat in excess of $10^{-8}$ \m\ yr$^{-1}$ which is still
very relevant in terms of central carbon ignition.  That value is much
less than the regime associated with the SSS.
The thermonuclear recurrent nova models do fall in the high mass
regime of the SSS accretion rate range.  There is at least
a suggestion, then, that if SSS can successfully accrete and
grow white dwarfs toward M$_{CH}$ that they may naturally
evolve through a phase when they are recurrent novae just
prior to their explosion as SN~Ia.

\section{Observational Constraints}
\label{cwconstraints}

Given the stubborn nature of the problem of identifying the progenitor systems
of SN~Ia, it is important to try to develop more observational tests that will
give direct constraints on various progenitor models.
Wheeler (1991, 1992)
discussed possible ways to constrain the
progenitors of SN~Ia by looking for evidence of various types of
circumstellar material from, e.g., mass transfer, an old common envelope,
a disrupted degenerate companion.  Cataclysmic variables of various
types, symbiotic variables, and the supersoft sources should
produce some hydrogen-rich circumstellar
matter from winds or sloppy mass transfer.  For the double degenerate model,
the nebula could be carbon and oxygen with very little hydrogen.

A quantitative attempt to set limits on some progenitor models in such a way
has been presented by Boffi and Branch (1995).
They note that some symbiotic variables represent wind accretion from a red
giant
onto a white dwarf.  The mass loss rate in this wind should be in the
range $10^{-7}$ to $10^{-5}$ \m\ yr$^{-1}$ which could, in principle,
be dense enough to generate observable radio
emission when shocked by supernova ejecta.  Boffi and Branch
and Eck \etal\ (1995) note that
the upper limits to radio emission may rule out such a symbiotic variable
progenitor for SN 1986G and possibly SN 1981B.  Other SN~Ia were not
observed in the radio at epochs that would provide useful constraints.
SN 1986G was a slightly peculiar event and the sample is in any case
very small, but this is a good illustration of a program that may provide
useful constraints.  These radio upper limits may also constrain the
 ``PNN mass loss" regulation mechanism for the SSS.

We have started a new program at McDonald Observatory to attempt to
routinely monitor the polarimetric characteristics of supernovae
(Wang \etal\ 1996).  By combining new observations with
the sparse data from the literature, Wang \etal\ identify a very
interesting trend.  All Type II supernovae seem to be polarized at a level
of about 1 percent and no Type Ia are polarized with upper limits of
about 0.2 percent.  This implies some substantial departure from spherical
symmetry for the SN II, although whether this is intrinsic to the
supernova ejecta in all cases or may also involve dust scattering
from a surrounding circumstellar nebula is not clear at this time.
Whatever the asymmetries, they may tell us something about the
rotational state or the presence of a binary companion.
Likewise, the limits on SN~Ia may imply a greater degree of spherical
symmetry, or the lack of a substantial circumstellar environment.
These upper limits of SN~Ia must be quantified, but hold some promise
for the constraint of progenitor models.  For instance, the double degenerate
model might be expected to begin with naturally large asymmetries.
The issue then is whether these asymmetries are wiped out in the post-explosion
expansion.  Alternatively, if the lack of polarization of SN~Ia implies
a low density circumstellar environment, the issue is whether the
polarization data provide any new constraints beyond the extant knowledge
that SN~Ia are not prodigious emitters of radio or X-ray radiation.
Qualitatively, the polarization could arise from electron scattering
in a photosphere which is geometrically distorted or asymmetrically
illuminated from within or by scattering from dust grains.  The latter
is much more efficient in imposing a polarization, so the limits
on the polarization of SN~Ia might in turn provide limits on the dust
density and distribution.  This might not give a useful constraint
on systems such as double degenerates which are not necessarily expected
to produce dust, but there might be useful constraints on progenitors systems
with cool, red, and hence potential dust forming, companions.  This might
include cataclysmic variables, symbiotic systems, or supersoft sources
with sub-giant companions.

An observation with great potential to constrain progenitor systems was
that of SN 1990M by Polcaro and Viotti (1991).
This paper reported a transient, narrow H$\alpha$ absorption line.  The
significance of this observation for constraining the state of a
hydrogen-rich circumstellar nebula and the difficulties it would raise
for the double-degenerate model was presented in Wheeler (1992).
Tsiopa (1995)
has constructed a plausible model for this feature
in terms of ``rings" of matter much like those observed in SN 1987A,
but smaller.  These rings can provide some absorption when viewed at the
right angle until swept up by the expanding ejecta.
This data has, however, been recently re-analyzed by Della Valle and
Benetti (1996).  It may be that the absorption feature reported by Polcaro
and Viotti was a result of incorrect galaxy subtraction.  Branch \etal\
(1983)
presented weak evidence for a transient
hydrogen emission line in SN 1981B.  This did not seem the result
of the reduction process, but was difficult to verify and does not
seem to have been reproduced in subsequent bright SN~Ia.  It is still
very important to continue high resolution observations of SN~Ia
that span their range of diversity to see if there are any
transient features associated with a circumstellar medium.

Meikle \etal\ (1995)
have reported the detection of He I $\lambda10830$
in SN 1994D.  If this is the correct identification, then the
velocity of the helium was rather low to be an outer skin of
helium, about 9,500 km s$^{-1}$, but rather high to be associated
with circumstellar matter.  Wheeler \etal\ (1995)
suggested the feature might rather be Mg II $\lambda$10926.  Meikle \etal\
conclude that He I provides a better fit than Mg II and suggest
that the helium may be that which results from a ``helium-rich"
freezeout in the ejecta.  If this interpretation is correct,
then the helium may have nothing to do with any circumstellar
matter, but may still represent a powerful constraint on
the nature of the progenitor (e.g.. Chandrasekhar versus sub-Chandrasekhar
mass) and hence indirectly on the question of progenitor evolution.
These observations clearly point up the vivid need for more infrared
spectroscopic observations of supernovae of all types.

One can also examine young supernova remnants for clues to
the circumstellar environment.  Some remnants are dominated by
Balmer emission with narrow, low velocity hydrogen lines.  These
are interpreted as arising from charge exchange between the
high velocity ejecta material and the low velocity neutral hydrogen that
is swept up in a collisionless shock.  This interpretation has
been applied to Tycho (SN 1572; Chevalier, Kirshner, and Raymond 1980)
and similar narrow hydrogen lines have recently been detected
in the spectrum of 3C58 (SN 1185; Smith 1995).
Although the status of the Balmer-dominated remnants is debated, there
are some grounds for identifying them as remnants of SN~Ia.  In any
case, the evidence for {\em neutral} hydrogen around the
remnants discourages their identification with progenitors which
had been SSS with extensive ionized nebulae (Kirshner, 1995).
This may not rule out a connection with SSS, since direct observations
only show one in nine SSS to have a distinct nebula (Remillard,
Rappaport and Macri 1995).  This constraint
may still help to develop a picture of the density of the ISM
associated with SN~Ia and perhaps the mass of any white dwarf, all
grist for the mill of determining the progenitor evolution.

\section{Conclusions}
\label{cwconcl}

The gathering evidence of the nature of SN~Ia shows that they do have a true
dispersion in photometric and spectral evolution.  The properties
of SN~Ia tend to correlate with the age of the stellar system.
There is no doubt that the weight of the evidence favors the
interpretation that SN~Ia are the thermonuclear explosion of
white dwarfs.  While the subject deserves (and will certainly
receive!) more study, the data, including the observed dispersion
in properties, is currently much more consistent
with the explosion of Chandrasekhar mass rather than sub-Chandrasekhar
mass models.  While there is still not an iota of direct evidence,
binary evolution is the only plausible way to lead to such an
explosion.

A key task is to concentrate on both theoretical and observational
means to discriminate possible binary star progenitors.  Ultimately,
this can not be too subtle a task.  Although estimates of mean rates of
production
of SN~Ia have declined a bit in recent years, SN~Ia are still a relatively
common phenomenon in galaxies like our own and there must be ample
progenitor systems if we only become wise or skilled
enough to recognize them.  Cataclysmic variables and novae are still
unlikely progenitors primarily because of the mass reservoir problem.
They still deserve more study in terms of issues like hibernation and the
multi-dimensional nature of nova hydrodynamics.  Symbiotic variables
also seem unlikely at this juncture because they are too few and
perhaps would be expected to be too radio loud.  The supersoft sources
are an exciting possibility, but still have problems.  Are there
enough of them?  Will they over-ionize their surroundings?  Double
degenerates seem unlikely to explain all SN~Ia, but they might
be needed to account for the odd, bright systems like SN 1991T.

The issue of Chandrasekhar mass versus sub-Chandrasekhar mass
progenitors is critical to this discussion.  It is, in principle,
much easier to provide the requisite number of sub-Chandrasekhar mass
progenitors. Nature, however, seems not to have chosen this route.
While we concentrate on the evolution that can lead to observed
SN~Ia, it might also be useful to contemplate the implications
for progenitor evolution of the seeming absence of sub-Chandrasekhar
mass explosions.  Taken literally, this seems to imply that evolving
binaries {\em avoid} conditions where they would spend substantial
periods accreting at rates of order $10^{-8}$ to
$10^{-7}$ \m\ yr$^{-1}$ where one expects detonation of an outer
degenerate helium layer.  It would be good to understand why
binary evolution avoids such conditions as a complement to
understanding what conditions are favored.

One of the most exciting recent developments in the study of SN~Ia
are the new {\em observational} initiatives to determine or at least
constrain the progenitors systems.  Much more thought needs to
be given to such efforts.

In the meantime, the quest for the binary progenitors of SN~Ia
continues.

\acknowledgements I am especially grateful for my education
in the topics covered
here to Zalman Barkat, Lars Bildsten, David Branch, Robert Harkness, Peter
H\"oflich,
Alexei Khokhlov, Mario Livio, Eli Livne,
Jim Truran and Lifan Wang.  I appreciate the support of the organizers of
the meeting for the chance to return to Cambridge despite
violating the Pringle age-limit theorem, for an entertaining, if odd,
game of cricket, and for teaching me even more wonderful things
you can do with a banana.
This research is supported in part by NSF Grant AST 9218035 and NASA
Grants NAGW-2905 and NAG5-2888.


\begin{thereferences}{}

\bibitem[]{}
Arnett, W. D., Branch, D., and Wheeler, J. C.: 1985,
\newblock {\em Nature} {\bf 314}, 337

\bibitem[]{}
Arnett, W. D., and Livne. E.: 1994,
\newblock {\em Astrophys. J.} {\bf 427}, 314

\bibitem[]{}
Barkat, Z. and Tuchman, Y.: 1980a,
\newblock {\em Astrophys. J.} {\bf 237}, 105

\bibitem[]{}
Barkat, Z. and Tuchman, Y.: 1980b,
\newblock {\em Astrophys. J.} {\bf 242}, 199

\bibitem[]{bartunov_arms}
Bartunov, O. S., Tsvetkov, D. Yu., and Filimonova, I. V. 1994:
\newblock {\em  Publ. Astron. Soc. Pac.} {\bf 106}, 1276

\bibitem[]{}
Biermann, P., and Tinsley, B. M.: 1974
\newblock {\em  Publ. Astron. Soc. Pac.} {\bf 86}, 791

\bibitem[]{bildsten_93}
Bildsten, L.: 1993,
\newblock {\em Astrophys. J.} {\bf 418}, L21

\bibitem[]{bildsten_95}
Bildsten, L.: 1995,
\newblock {\em Astrophys. J.} {\bf 438}, 852

\bibitem[]{boffi_branch}
Boffi, and Branch, D.: 1995,
\newblock {\em  Publ. Astron. Soc. Pac.} {\bf 107}, 347

\bibitem[]{boisseau_thesis}
Boisseau, J. R.: 1995, PhD thesis, University of Texas at Austin

\bibitem[]{}
Bragaglia, A., Greggio, L., Renzini, A., and D'Odorico, S.:1990,
\newblock {\em Astrophys. J.} {\bf 365}, L13

\bibitem[]{branch_81}
Branch, D.: 1981,
\newblock {\em Astrophys. J.} {\bf 248}, 1076

\bibitem[]{branch_81b}
Branch, D., Lacy, C. H., McCall, M. L., Sutherland, P. G., Uomoto, A.,
Wheeler, J. C., and Wills, B. J. 1983,
\newblock{\em Astrophys. J.} {\bf 270}, 123

\bibitem[]{branch_progen}
Branch, D., Livio, M., Yungleson, L. R., Boffi, F. R., and Baron, E.: 1995,
\newblock {\em Publ. Astron. Soc. Pac.} {\bf XXX}, xxx

\bibitem[]{branch_vdb}
Branch, D. and van den Bergh, S.: 1993,
\newblock {\em, Astron. J.}, {\bf 105}, 2231

\bibitem[]{CKR_tycho}
Chevalier, R. A., Kirshner, R. P. and Raymond, J. C.: 1980,
\newblock {\em Astrophys. J.} {\bf 235}, 186

\bibitem[]{della_arms}
Della Valle, M., and Benetti, S.: 1996,
\newblock {\em Astron. Astrophys.} {\bf XXX}, xxx

\bibitem[]{della_arms}
Della Valle, M., and Livio, M.: 1994,
\newblock {\em Astrophys. J.} {\bf 423}, L31

\bibitem[]{}
Di Stefano, R. and Rappaport, S.: 1994, 
\newblock {\em  Astrophys. J.}, {\bf 437}, 733.

\bibitem[]{}
Eck, C. R., Cowan, J. J., Roberts, D. A., Boffi, F. R., and Branch, D.: 1995,
\newblock {\em  Astrophys. J.}, {\bf xxx}, Lxx

\bibitem[]{filipp91T}
Filippenko, A. V. \etal: 1992a,
\newblock {\em  Astrophys. J.}, {\bf 384}, L15

\bibitem[]{filipp91bg}
Filippenko, A. V. \etal: 1992b,
\newblock {\em, Astron. J.}, {\bf 104}, 1543

\bibitem[]{fisher_calcium}
Fisher, A., Branch, D., H\"oflich, P., and Khokhlov, A.: 1995,
\newblock {\em Astrophys. J.} {\bf 447}, L73

\bibitem[]{}
Foss, D., Wade, R. A., and Green, R. F.: 1991,
\newblock {\em Astrophys. J.} {\bf 374}, 281

\bibitem[]{hamuy94}
Hamuy, M., Phillips, M. M., and Maza, J.: 1994
\newblock {\em Astron. J.} {\bf 108}, 2226

\bibitem[]{hamuy96_theory}
Hamuy, M., H\"oflich, P., Khokhlov, A., Phillips, M. M.,
Suntzeff, N., and Wheeler, J. C.: 1996b,
\newblock {\em Astrophys. J.} {\bf XXX}, xxx

\bibitem[]{hamuy95_obs} 
Hamuy, M., Phillips, M. M., Maza, M., Suntzeff, N., Schommer, R. A.,
Aviles, A.: 1995
\newblock {\em Astron. J.} {\bf 109}, 1

\bibitem[]{hamuy96_obs} 
Hamuy, M., Phillips, M. M., Maza, M., Suntzeff, N., Schommer, R. A.,
Aviles, A.: 1996a
\newblock {\em Astrophys. J.} {\bf XXX}, xxx

\bibitem[]{}
Harkness, R. P.: 1986,
\newblock in D. Mihalas and K.-H.A. Winkler (eds.), {\em Stars and Compact
Objects}, 166,
Springer-Verlag, Berlin

\bibitem[]{}
Harkness, R. P.: 1991a,
\newblock in S. E. Woosley (ed.), {\em Supernovae}, 454, Springer-Verlag, New
York

\bibitem[]{}
Harkness, R. P.: 1991b,
\newblock in {\em SN1987A and Other Supernovae}, J. Danziger and K. Kj\"ar
(eds.),
447, ESO, Garching

\bibitem[]{}
Hasinger, G.: 1994,
\newblock in {\em The Evolution of X-Ray Binaries}, S. S. Holt and C. Day
(eds.),
611, AIP Press, New York

\bibitem[]{hoflich_94D}
H\"oflich, P.: 1995, 
\newblock {\em Astrophys. J.} {\bf 443}, 89

\bibitem[]{HKW95}
H\"oflich,P., Khokhlov,A., and Wheeler, J. C.: 1995,
\newblock {Astrophys. J.} {\bf 444}, 831

\bibitem[]{HK96}
H\"oflich,P. and Khokhlov,A.: 1996,
\newblock {\em Astrophys. J.} {\bf XXX}, xxx

\bibitem[]{HK95}
Howell, S. B., Szkody, P., and Cannizzo, J. K.: 1995,
\newblock {\em Astrophys. J.} {\bf 439}, 337

\bibitem[]{}
Iben, I. Jr., and Tutukov, A. V.: 1991
\newblock in D. L. Lambert (ed.),
{\em Proceedings of the McDonald Observatory Fiftieth Anniversary Symposium,
Frontiers of Stellar Evolution}, 403,
Astronomical Society of the Pacific, Provo

\bibitem[]{}
Iben, I., Jr., and Tutukov, A. V.: 1984,
\newblock {\em Ap. J. Suppl.} {\bf 54}, 335

\bibitem[]{kenyon_symb}
Kenyon, S. J., Livio, M., Mikolajewska, J., and Tout, C.: 1993,
\newblock {\em Astrophys. J.} {\bf 407}, L81

\bibitem[]{khokhlov_delaydet}
Khokhlov, A.: 1991a,
\newblock {\em Astron. Astrophys.} {\bf 245}, 114

\bibitem[]{khokhlov_PDD}
Khokhlov, A.: 1991b,
\newblock {\em Astron. Astrophys.} {\bf 245}, L25

\bibitem[]{khokhlov_detstab}
Khokhlov, A.: 1993,
\newblock {\em Astrophys. J.} {\bf 419}, 200

\bibitem[]{khokhlov_3d}
Khokhlov, A.: 1995,
\newblock {\em Astrophys. J.} {\bf 449}, 695

\bibitem[]{KMH93_PDD}
Khokhlov, A. M\"uller, E. and H\"oflich, P.: 1993,
\newblock {\em Astron. Astrophys.} {\bf 270}, 223

\bibitem[]{KOWI}
Khokhlov, A., Oran, E. R., and Wheeler, J. C.: 1995
\newblock {\em Combust. and Flame} {\bf XXX}, xxx

\bibitem[]{KOWII}
Khokhlov, A., Oran, E. R., and Wheeler, J. C.: 1996,
\newblock {\em Combust. and Flame} {\bf XXX}, xxx

\bibitem[]{kirshner_pc}
Kirshner, R. P.: 1995,
\newblock in R. Canal, P. Ruiz-Lapuente (eds.),
{\em NATO ASI on Thermonuclear Supernovae}, in press, Kluwer Academic
Publishers, New York

\bibitem[]{leib91bg}
Leibundgut \etal: 1993, {\em PASP} {\bf 107}, 347

\bibitem[]{livio_magnova}
Livio, M.: 1983,
\newblock {\em Astron Astrophys.} {\bf 121}, L7

\bibitem[]{livio_novacyg}
Livio, M., Shankar, A., and Truran, J. W:. 1988,
\newblock {\em Astrophys. J.} {\bf 330}, 264

\bibitem[]{livio_commonenv}
Livio, M., Shankar, A., Burkert, A. and Truran, J. W.: 1990,
\newblock {\em Astrophys. J.} {\bf 356}, 250

\bibitem[]{livio_truran}
Livio, M. and Truran, J. W. 1992,
\newblock {\em Astrophys. J.} {\bf 389}, 695

\bibitem[]{livne_nova}
Livne, E.: 1995,
\newblock in R. Canal, P. Ruiz-Lapuente (eds.),
{\em NATO ASI on Thermonuclear Supernovae} in press, Kluwer Academic
Publishers, New York

\bibitem[]{livne_subCh}
Livne, E., and Arnett, W. D.: 1995
\newblock {\em Astrophys. J.} {\bf 452}, 62

\bibitem[]{livne_subCh}
Livne, E., and Glasner, A. S.: 1991
\newblock {\em Astrophys. J.} {\bf 370}, 272

\bibitem[]{long_etalSSS}
Long, K. S., Helfand, D. J., and Grebelsky D. A.: 1981
\newblock {\em Astrophys. J.} {\bf 248}, 925

\bibitem[]{marsh_dd}
Marsh, T. R., Dhillon, V. S., and Duck, S. R. 1995,
\newblock {\em  Mon. Not. R. Astron. Soc.} {\bf 275}, 828

\bibitem[]{}
Maza, J., Hamuy, M., Phillips, M. M., Suntzeff, N. B., and Ayiles, R.: 1994 %
\newblock {\em  Astrophys. J.} {\bf 424}, L107

\bibitem[]{maza_vdb}
Maza, J. and van den Bergh, S,: 1976
\newblock {\em  Astrophys. J.} {\bf 204}, 519

\bibitem[]{maza_vdb}
Macdonald, J.: 1986
\newblock {\em  Astrophys. J.} {\bf 305}, 251

\bibitem[]{mcm_ciard}
McMillan, R. and Ciardullo, R.: 1995,
\newblock in R. Canal, P. Ruiz-Lapuente (eds.),
{\em NATO ASI on Thermonuclear Supernovae} in press, Kluwer Academic
Publishers, New York

\bibitem[]{meikle_94D}
Meikle, W. P. S. \etal: 1995,
\newblock in preparation

\bibitem[]{munari_renzini_symb}
Munari, U., and Renzini, A.: 1992,
\newblock {\em  Astrophys. J.} {\bf 397}, L87

\bibitem[]{NTY84}
Nomoto, K., Thielemann, F.-K., and Yokoi, K.: 1984,
\newblock {\em Astrophys. J.} {\bf 286}, 644

\bibitem[]{oemler_tinsley}
Oemler, A. Jr. and Tinsley, B. M.: 1979
\newblock {\em Astron. J.} {\bf 84}, 985

\bibitem[]{}
Paczynski, B. E.: 1985,
\newblock in D. Q. Lamb and J. Patterson (eds.), {\em Cataclysmic Variables
and Low Mass X-ray Binaries}, 1, Reidel, Dordrecht

\bibitem[]{pakull_SSS}
Pakull, M. W., Moch, C., Bianchi, L., Thomas, H.-C., Guibert, J.,
Beaulieu, P., Grisom, P., and Schaeidt, S.: 1993,
\newblock {\em Astron. Ap.} {\bf 278}, L39

\bibitem[]{phillips93}
Phillips, M. M.: 1993,
\newblock {\em Astrophys. J.} {\bf 413}, L105

\bibitem[]{phillips91T}
Phillips, M. M., Wells, L. A., Suntzeff, N. B., Hamuy, M.,
Leibundgut, B., Kirshner, R. P.,
and Foltz, C. B.: 1992, {\em Astron. J.}, {\bf 103}, 1632

\bibitem[]{polcaro_viotti}
Polcaro, V. F. and Viotti, R.: 1991,
\newblock {\em Astron. Ap.} {\bf 242}, L9

\bibitem[]{pskovskii_77}
Pskovskii, Yu. P.: 1977
\newblock {\em Sov.Astron.} {\bf 21}, 675

\bibitem[]{}
Rappaport, S., Chiang, E., Kallman, T. and Malina, R.: 1994,
\newblock {\em Astrophys. J.} {\bf 431}, 237

\bibitem[]{}
Rappaport, S., DiStefano, R. and Smith, J. D.: 1994, 
\newblock {\em Astrophys. J.} {\bf 426}, 692.

\bibitem[]{}
Reiss, A. G., Press, W. H., and Kirshner, R. P.: 1995
\newblock {\em Astrophys. J.} {\bf 438}, L17

\bibitem[]{}
Remillard, R. A., Rappaport, S., and Macri, L. M.: 1995,
\newblock {\em Astrophys. J.} {\bf 439}, 646

\bibitem[]{robinson_shafter}
Robinson, E. L. and Shafter, A. W.: 1987,
\newblock {\em Astrophys. J.} {\bf 322}, 296

\bibitem[]{}
Ruiz-Lapuente, P., Burkert, A., and  Canal, R.,: 1995,
\newblock {\em Astrophys. J.} {\bf 447}, L1

\bibitem[]{SCW}
Shafter, A., Wheeler, J. C., and  Cannizzo, J.,: 1986,
\newblock {\em Astrophys. J.} {\bf 305}, 261

\bibitem[]{shara_dimcv}
Shara, M., Moffatt, A., Potter, M., Bode, M., and Stephenson, F. R.: 1993,
\newblock {\em Ann. Israel Phys. Soc.} {\bf 10}, 84

\bibitem[]{smith_3c58}
Smith, R. C.: 1995,
\newblock in R. Canal, P. Ruiz-Lapuente (eds.),
{\em NATO ASI on Thermonuclear Supernovae} in press, Kluwer Academic
Publishers, New York

\bibitem[]{}
Suntzeff, N.: 1995,
\newblock in R. Canal, P. Ruiz-Lapuente (eds.),
{\em NATO ASI on Thermonuclear Supernovae} in press, Kluwer Academic
Publishers, New York

\bibitem[]{}
Sutherland, P. G., and Wheeler, J. C.: 1984,
\newblock {\em Ap. J.} {\bf 280}, 282

\bibitem[]{trumperSSS}
Tr\"umper, J., Hasinger, G., Aschenback, B., Br\"auninger, H.,
Briel, U. G., Burkert, W., Fink, H., Pfefferman, E., Pietsch, W.,
Predehl, P., Schmitt, J. H. M. M., Voges, W., Zimmerman, U.,
and Beuermann, K.: 1991,
\newblock {\em Nature} {\bf 349}, 579

\bibitem[]{tsiopa}
Tsiopa, O.: 1995,
\newblock in R. Canal, P. Ruiz-Lapuente (eds.),
{\em NATO ASI on Thermonuclear Supernovae} in press, Kluwer Academic
Publishers, New York

\bibitem[]{vdb_IRrates}
van den Bergh, S.: 1990,
\newblock {\em  Publ. Astron. Soc. Pac.} {\bf 102}, 1318

\bibitem[]{vdh_SSS}
van den Heuvel, E. P. J., Bhattacharya, D., Nomoto, K., and Rappaport, S. A.:
1992,
\newblock {\em Astron. Astrophys.} {\bf 262}, 97

\bibitem[]{wangpol95}
Wang, L., Li, Z., Clocchiatti, A., and Wheeler, J. C.: 1996,
\newblock {\em Astrophys. J.} {\bf XXX}, xxx

\bibitem[]{warner_brake}
Warner, B.: 1991,
\newblock in D. L. Lambert (ed.),
{\em Proceedings of the McDonald Observatory Fiftieth Anniversary Symposium,
Frontiers of Stellar Evolution}, 451,
Astronomical Society of the Pacific, Provo

\bibitem[]{}
Webbink, R. F.: 1984,
\newblock {\em Ap. J. (Letters)} {\bf 277}, 355

\bibitem[]{}
Wells \etal: 1994,
\newblock {\em Astronom. J.} {\bf 108}, 2233

\bibitem[]{wheeler_mcd}
Wheeler, J. C.: 1991,
\newblock in D. L. Lambert (ed.),
{\em Proceedings of the McDonald Observatory Fiftieth Anniversary Symposium,
Frontiers of Stellar Evolution}, 483,
Astronomical Society of the Pacific, Provo

\bibitem[]{wheeler_cordoba}
Wheeler, J. C.: 1992,
\newblock {\em Evolutionary Processes in Interacting Binary Systems}, 225,
IAU Symp. \ 151, Kluwer, Dordrecht

\bibitem[]{wheeler_ANNREV}
Wheeler, J. C., Sneden, C., and Truran, J. W. Jr.: 1989,
\newblock {\em Ann. Rev. Astron. Ap.}  {\bf 27}, 279

\bibitem[]{wheeler_fowler}
Wheeler, J. C., Swartz, D., and Harkness, R. P.: 1993,
\newblock {\em Phys. Rprts.}  {\bf 227}, 113

\bibitem[]{woosley_subCh}
Woosley, S. E., and Weaver, T. A.: 1994,
\newblock {\em Astrophys. J.} {\bf 423}, 371

\bibitem[]{yungelson_SSS}
Yungelson, L., Livio, M., Truran, J. W., Tutukov, A., and Federova, A.: 1996
\newblock {\em Astrophys. J.} {\bf XXX}, xxx

\bibitem[]{}
Zel'dovich, Ya. B., Librovich, V. B., Makhviladze, G. M., and Sivashinksy, G.
I.: 1969,
\newblock in {\em 2nd Intern. Colloq. on Explosion and Reacting Systems
Gasdynamics} ,
\ 10, Novosibirsk

\bibitem[]{}
Zel'dovich, Ya. B., Librovich, V. B., Makhviladze, G. M., and Sivashinksy, G.
I.: 1970,
\newblock {\em Astronaut. Act.}, {\bf 15}, 313


\end{thereferences}
\end{document}